\documentstyle[eqsecnum,aps,epsbox]{revtex}

\begin{document}
\draft
\title{{\bf CP Violation in Neutrinoless Double Beta Decay \\
and Neutrino Oscillation}}

\author{T. FUKUYAMA and K. MATSUDA}
\address{Department of Physics, \\
        Ritsumeikan University, Kusatsu, \\
        Shiga, 525 Japan}
\author{H. NISHIURA}
\address{Department of General Education, \\
        Junior College of Osaka Institute of Technology, \\
        Asahi-ku, Osaka,535 Japan}

\date{November 13, 1997}
\maketitle



\begin{abstract}
Taking possible CP violations into account, we discuss constraints 
on lepton mixing angles obtained from neutrinoless double beta decay 
and from neutrino oscillation for three flavour Majorana neutrinos. 
From the CHORUS oscillation experiment, 
combined with the data from neutrinoless double beta decay, 
we show that the large angle solution for \(\theta_{23}\) is improbable 
if the neutrino mass \(m_3\) of the third generation is 
a candidate for hot dark matter.
\end{abstract}




\narrowtext
\section{Introduction}
It is one of the most important problems in particle physics 
whether neutrinos have mass or not. 
From recent neutrino experiments 
\cite{Kamioka} \cite{Homestake} \cite{Gallex} \cite{Sage},
it becomes very probable 
that
neutrinos have mass. However, if neutrinos have mass, 
we must explain the 
reason why it would be so small relative to charged lepton masses. 
The seesaw mechanism 
is one of the most promising candidates for such an
explanation. In this case, neutrinos become 
Majorana particles, and three CP violating phases remain 
in the lepton mixing matrix U for the three generation case \cite{bilenky}. 
This contrasts with the case of Dirac Neutrinos, 
in which only one CP-violating  phase remains in U. Thus, 
for Majorana neutrinos, possible CP violations due to these phases 
in U complicates obtaining information 
about the lepton mixing angles from the corresponding experiments.

In this paper we consider neutrinos as massive Majorana particles 
with three generations and shaw how experiments, 
such as neutrinoless double beta decay 
(\((\beta \beta)_{0 \nu}\)) and neutrino oscillation, 
constrain lepton mixing angles and neutrino mass. 
In order to obtain these constraints 
we take possible CP-violating phases in U into account. 
In \((\beta \beta)_{0 \nu}\) there are 
three CP-violating phases which contribute to possible CP violations. 
In the neutrino oscillation, on the other hand, 
there is only one such phase as in the case of Dirac neutrinos.

In sec.II
we obtain  constraints from \((\beta \beta)_{0 \nu}\).
The constraint which arises from the neutrino oscillation experiment CHORUS is 
described in sec.III. Combining these constraints, we find that the large mixing 
angle of \(\theta_{23}\) becomes 
improbable if the third generation of neutrino is 
a candidate for hot dark matter and not degenerate.
\par

\section{Neutrinoless double beta decay \label{sec2}}

Let us first consider the neutrinoless double beta decay 
($(\beta\beta)_{0{\nu}}$) which occurs only in the case of Majorana
neutrinos.  
\begin{center}
 --- FIG.1 ---
\end{center}
The decay rate of $(\beta\beta)_{0{\nu}}$ is, 
in the absence of right-handed couplings, 
proportional to the "averaged" mass defined by \cite{doi}
\begin{equation} 
\langle m_{\nu} \rangle\equiv |\sum _{j=1}^{3}U_{ej}^2m_j|.\label{eq1}
\end{equation}
Here $U_{\alpha j}$ is the left-handed neutrino mixing matrix 
which combines the weak eigenstate ($\alpha =e,\mu$ and $\tau$) 
to the mass eigenstate with mass \(m_j\) (j=1,2 and 3). 
$U_{\alpha j}$ takes the following form in the case of Majorana neutrinos,
\widetext
\begin{equation}
U=
\left(
\begin{array}{ccc}
c_1c_3&s_1c_3e^{i\beta}&s_3e^{i(\rho-\phi )}\\
(-s_1c_2-c_1s_2s_3e^{i\phi})e^{-i\beta}&c_1c_2-s_1s_2s_3e^{i\phi}&s_
2c_3e^{i(\rho-\beta )}\\
(s_1s_2-c_1c_2s_3e^{i\phi})e^{-i\rho}&(-c_1s_2-s_1c_2s_3e^{i\phi})e^
{-i(\rho-\beta )}&c_2c_3\\
\end{array}
\right).\label{eq2}
\end{equation}
\narrowtext
Here $c_j=cos\theta_j$, $s_j=sin\theta_j$ 
($\theta_1=\theta_{12},~\theta_2=\theta_{23},~\theta_3=\theta_{31}$).  
Besides $\phi$ there appear two additional CP violating phases 
$\beta$ and $\rho$ for Majorana neutrinos.  
Hence $\langle m_{\nu} \rangle$ becomes 
\begin{equation}
\langle m_{\nu} \rangle=|m_1c_1^2c_3^2-m_2s_1^2c_3^2e^{-2i\beta 
'}-m_3s_3^2e^{-2i\rho '}|,\label{eq3}
\end{equation}
where we have introduced 
\begin{equation}
\beta '\equiv \frac{\pi}{2}-\beta ,\quad \rho'
\equiv \frac{\pi}{2}-(\rho-\phi).\label{eq4}
\end{equation}
The CP violating phases \(\beta'\) and \(\rho'\) complicate 
extracting the constraints 
on the mixing angles from \(\langle m_{\nu} \rangle\). 
In the following discussion, 
we follow the method given in \cite{nishiura} 
in order to get the constraints which are independent on \(\beta'\) and 
\(\rho'\).
\par
From Eq.(\ref{eq3}) it follows that
\begin{eqnarray}
\langle m_{\nu} \rangle^2 &=& (m_1c_1^2c_3^2-m_2s_1^2c_3^2cos2\beta 
'-m_3s_3^2cos2\rho ')^2 \nonumber\\
&& +(m_2s_1^2c_3^2sin2\beta '+m_3s_3^2sin2\rho ')^2 \label{eq5}
\end{eqnarray}
Rewriting $cos2\rho '$ and $sin2\rho '$ as $tan\rho '$, we can 
consider
Eq.(\ref{eq5}) 
as an equation of $tan\rho '$:
\begin{equation}
a_{+\beta}tan^2\rho '+b_{\beta}tan\rho '+a_{-\beta}=0. \label{eq6}
\end{equation}  
Here $a_{\pm \beta}$ and $b_{\beta}$ are defined by
\begin{eqnarray}
a_{\pm \beta}&\equiv& 4sin^2\beta 'm_2s_1^2c_3^2
(m_1c_1^2c_3^2 \pm m_3s_3^2) \nonumber\\
&& +(m_1c_1^2c_3^2-m_2s_1^2c_3^2\pm m_3s_3^2)^2-
\langle m_{\nu} \rangle^2 \label{eq7} \\
b_\beta&\equiv& 4m_2m_3s_1^2s_3^2c_3^2sin2\beta '.\nonumber
\end{eqnarray}
So, the discriminant $D$ for 
Eq.(\ref{eq6})
must satisfy the following inequality:
\begin{eqnarray}
D&\equiv& b_\beta^2-4a_{+\beta}a_{-\beta}\nonumber\\
&=&4^3(m_1c_1^2c_3^2)^2(m_2s_1^2c_3^2)^2
(f_+-sin^2\beta')(sin^2\beta'-f_-) \nonumber\\
&\geq& 0,\label{eq8}
\end{eqnarray}
where 
\begin{equation}
f_{\pm}\equiv \frac{(\langle m_{\nu} \rangle\pm 
m_3s_3^2)^2-(m_1c_1^2c_3^2-m_2s_1^2c_3^2)^2}
{4m_1m_2c_1^2s_1^2c_3^4}. \label{eq9}
\end{equation}
So we obtain
\begin{equation}
f_-\leq sin^2\beta '\leq f_+.\label{eq10}
\end{equation}
It follows from Eq.(\ref{eq10}) that
\begin{equation}
f_-\leq 1,~~ f_+\geq 0. \label{eq11}
\end{equation}
Analogously, rewriting $cos2\beta '$ and $sin2\beta '$ as 
$tan\beta'$, and considering Eq.(\ref{eq5}) 
as an equation of $tan\beta '$, we obtain the inequalities:
\begin{equation}
g_-\leq sin^2\rho '\leq g_+. \label{eq12}
\end{equation}
Here
\begin{equation}
g_{\pm}\equiv \frac{(\langle m_{\nu} \rangle\pm 
m_2s_1^2c_3^2)^2-(m_1c_1^2c_3^2-m_3s_3^2)^2}{
4m_1m_3c_1^2s_3^2c_3^2}. \label{eq13}
\end{equation}
So we get 
\begin{equation}
g_-\leq 1,~~ g_+\geq 0. \label{eq14}
\end{equation}
Conditions (\ref{eq11}) and (\ref{eq14}) are consistency conditions. 
The CP violation area is given by more stringent conditions
\begin{equation}
0\leq f_-\leq sin^2\beta '\leq f_+\leq 1 \
or \
0\leq g_-\leq sin^2\rho '\leq g_+\leq 1. \label{eq15}
\end{equation}
Using the inequalities (\ref{eq11}) and (\ref{eq14}), 
we can describe the allowed region for mixing 
angles in the $s_1^2$ versus $s_3^2$ plane 
once the neutrino masses $m_i$ and the 
"averaged " neutrino  mass $\langle m_{\nu}\rangle $ are known.
Only an upper bound of O(1eV) 
for the magnitude of \(\langle m_\nu \rangle\) has been determined 
experimentally \cite{balysh}.
\par
The neutrino masses may be safely ordered as \(m_1 \le m_2 \le m_3\).
So in the following discussions we will consider three cases:

a) $\langle m_\nu \rangle \leq m_1$,
\par

b) $m_1 \leq \langle m_\nu \rangle \leq m_2$,
\par
and
\par
c) $m_2 \leq \langle m_\nu \rangle \leq m_3$.

Note that the definition of $\langle m_{\nu}\rangle $ in Eq.(\ref{eq1}) 
and the Schwartz inequality jointly imply that
\begin{equation}
\langle m_\nu \rangle \leq \sum _{j=1}^{3} m_j|U_{ej}^2|\leq m_3\sum
_{j=1}^{3} 
|U_{ej}^2|=m_3, \label{eq16}
\end{equation}
that is, $\langle m_\nu \rangle$ can not be larger than $m_3$.

The allowed regions in the 
$s_1^2$ versus $s_3^2$ plane 
for cases (a), (b) and (c) are obtained from 
Eqs.(\ref{eq11}) and (\ref{eq14}), and are shown in FIG.2.
\begin{center}
--- FIG.2 ---
\end{center}
From FIG.2, we obtain the following upper bound on $s_3^2$ :
\begin{equation}
s_3^2\leq \frac{m_2+\langle m_{\nu} \rangle}{m_3+m_2} \label{eq17}
\end{equation}
in each of these cases.
The CP violation areas in the $s_1^2$ versus $s_3^2$ plane given 
by Eq.(\ref{eq15}) are indicated by oblique lines in FIG.2.
Case (a) has been considered also in \cite{nishiura} 
and\cite{minakata}.  In \cite{nishiura}, the 
representation for the mixing matrix adopted by 
Cabibbo-Kobayashi-Maskawa was used. In \cite{minakata}, only 
the limiting case 
where all the neutrino masses are degenerate ($m_1=m_2=m_3$) 
was discussed.  
It should be noted that we consider the additional cases (b) and (c) and 
also that no conditions on the neutrino masses have been imposed so far.

FIG.2 describes the allowed regions for neutrinoless double beta decay 
in the most general case. 
In the following discussion we restrict cases (b) and (c) as follows: 
(b\('\)) 
\(m_1 \ll \langle m_\nu \rangle < m_2 \sim m_3\) 
and (c\('\)) \(m_1 \sim m_2 \ll \langle m_\nu \rangle < m_3\). We will see how the 
neutrinoless double beta decay constrains the mixing angles 
\(\theta_2\) and \(\theta_3\). 

In case (b\('\)) we obtain 
\(\frac{m_1+\langle m_\nu \rangle}{m_3+m_1} \approx 
\frac{\langle m_\nu \rangle-m_1}{m_3-m_1} \approx 
\frac{\langle m_\nu \rangle}{m_3}\),
\(\frac{\langle m_\nu \rangle-m_1}{m_2-m_1} \approx 
\frac{m_1+\langle m_\nu \rangle}{m_2+m_1} \approx 
\frac{\langle m_\nu \rangle}{m_3}\) 
and 
\(\frac{m_2 \pm \langle m_\nu \rangle}{m_3+m_2} \approx 
\frac{1}{2} \pm \frac{\langle m_\nu \rangle}{2m_3}\). 
So the allowed regions depend solely on the value of 
\(\frac{\langle m_\nu \rangle}{m_3}\). 
In FIG.3, we depict the allowed region 
 \(s_1^2\) versus \(s_3^2\) plane for 
\(\frac{\langle m_\nu \rangle}{m_3}=\)
0.16, 0.33, 0.49, 0.65, 0.82 \ and \ 0.98.

\begin{center}
--- FIG.3 ---
\end{center}

In case (c\('\)) we have 
\(
\frac{m_1 + \langle m_\nu \rangle}{m_3+m_1} \approx 
\frac{m_2 + \langle m_\nu \rangle}{m_3+m_2} \approx
\frac{\langle m_\nu \rangle - m_2}{m_3-m_2} \approx 
\frac{\langle m_\nu \rangle}{m_3}
\) .
Therefore, the allowed region is 
\(s_3^2 \approx \frac{\langle m_\nu \rangle}{m_3}\),
\(0 \le s_1^2 \le 1\).
It is interesting that the value of \(s_3^2\) is completely determined by 
\(\frac{\langle m_\nu \rangle}{m_3}\) in this case.

\section{Neutrino oscillation \label{sec3}}

In this section, we consider the constraints obtained 
from the neutrino oscillation 
experiment CHORUS\cite{chorus} and see how the method 
developed in the previous section can also be used to describe 
the allowed region in $s_1^2$ versus $s_3^2$ plane.

The CHORUS experiment investigates \(\nu_\mu \to \nu_\tau\) 
oscillation through the observation of the \(\tau\) leptons.
The \(90 \% \) C.L. upper limit of the probability of 
\(P(\nu_\mu \to \nu_\tau)\) is \(2.5 \times 10^{-3}\) \cite{shibuya}.
In order to obtain the constraints 
from this experiment we consider two cases:

\begin{description}
	\item[(i)] \(m_3 \sim m_2 \gg m_1\)
	\item[(ii)] \(m_3 \gg m_2 \sim m_1\)
\end{description}

Firstly we consider case (i). In this case we have
\(\delta m_{31}^2 \equiv m_3^2-m_1^2 \sim 
\delta  m_{21}^2 \equiv m_2^2-m_1^2 \gg 
\delta m_{32}^2 \equiv m_3^2-m_2^2\). 
Substituting the experimental setting, \(E_\nu = 27\)GeV and 
\(L \approx 0.6 \mbox{km}\), we have 
\(\frac{\delta m^2 L}{4E_\nu} = 
2.8 \times 10^{-2} \delta m^2 \, \mbox{eV}^2.\)
Hence we may set 
\(sin \left( \frac{\delta m_{32}^2 L}{4E_\nu} \right) \sim 0\). So the 
approximate oscillation probability is given by \cite{tanimoto}
\begin{equation}
P(\nu_{\mu}\rightarrow\nu_{\tau})=4|U_{\mu 1}|^2|U_{\tau 1}|^2sin^2
(\frac{\delta m_{31}^2L}{4E_\nu}). \label{eq18}
\end{equation}
Substituting the expression of Eq.(\ref{eq2}) into Eq.(\ref{eq18}),
we obtain the following equation w.r.t \(\cos \phi\),
\begin{eqnarray}
\frac{P(\nu_\mu\rightarrow\nu_\tau)}
{4\sin^2(\frac{\delta m_{31}^2}{4E_\nu}L)}
& = & a_+ \cos^2\phi-2b\cos\phi+a_- \label{eq19}\\
& \equiv & f(\cos\phi).\nonumber
\end{eqnarray}
Here
\begin{eqnarray}
 a_+&\equiv &-(2s_1s_2s_3c_1c_2)^2,\nonumber\\
 a_-&\equiv 
&(s_1^2c_2^2+c_1^2s_2^2s_3^2)(s_1^2s_2^2+c_1^2c_2^2s_3^2), \label{eq20}\\
 b&\equiv 
 &s_1s_2s_3c_1c_2(s_1^2-c_1^2s_3^2)(c_2^2-s_2^2).\nonumber
\end{eqnarray}
The oscillation process does not distinguish 
Majorana neutrino from Dirac neutrino 
and only the $\phi$ phase occurs. 

The constraints are obtained from Eq.(\ref{eq19}). \(a_+\) 
is negative definite 
and 
\(f(\pm1)\)
is positive definite. Therefore the condition 
$-1\leq cos\phi \leq 1$ implies the following inequalities:
\widetext
\begin{description}
\item[\qquad \qquad Case a-1 : ] 
\(
 0 \le \frac{(s_2^2-c_2^2)(s_1^2-c_1^2s_3^2)}{4s_1s_2s_3c_1c_2} \le 1
\)
\begin{equation}
(s_1c_2-c_1s_2s_3)^2(s_1s_2+c_1c_2s_3)^2 \le 
\frac{P(\nu_\mu\rightarrow\nu_\tau)}{4\sin^2(\frac{\delta
m_{31}^2}{4E_\nu}L)} 
      \le \frac{1}{4}(s_1^2+c_1^2s_3^2)^2 \label{eq21}
\end{equation}
\item[\qquad \qquad Case a-2 : ]
\(
1<\frac{(s_2^2-c_2^2)(s_1^2-c_1^2s_3^2)}{4s_1s_2s_3c_1c_2}
\)
\begin{equation}
(s_1c_2-c_1s_2s_3)^2(s_1s_2+c_1c_2s_3)^2\le
\frac{P(\nu_\mu\rightarrow\nu_\tau)}{4\sin^2(\frac{\delta
m_{31}^2}{4E_\nu}L)}
\le (s_1c_2+c_1s_2s_3)^2(s_1s_2-c_1c_2s_3)^2 \label{eq22}
\end{equation}
\item[\qquad \qquad Case b-1 : ]
\(
-1 \le \frac{(s_2^2-c_2^2)(s_1^2-c_1^2s_3^2)}{4s_1s_2s_3c_1c_2} \le 0
\)
\begin{equation}
(s_1c_2+c_1s_2s_3)^2(s_1s_2-c_1c_2s_3)^2 \le 
\frac{P(\nu_\mu\rightarrow\nu_\tau)}{4\sin^2(\frac{\delta
m_{31}^2}{4E_\nu}L)} 
      \le \frac{1}{4}(s_1^2+c_1^2s_3^2)^2 \label{eq23}
\end{equation}
\item[\qquad \qquad Case b-2 : ]
\(
\frac{(s_2^2-c_2^2)(s_1^2-c_1^2s_3^2)}{4s_1s_2s_3c_1c_2}<-1
\)
\begin{equation}
(s_1c_2+c_1s_2s_3)^2(s_1s_2-c_1c_2s_3)^2\le
\frac{P(\nu_\mu\rightarrow\nu_\tau)}{4\sin^2(\frac{\delta
m_{31}^2}{4E_\nu}L)}
\le (s_1c_2-c_1s_2s_3)^2(s_1s_2+c_1c_2s_3)^2 \label{eq24}
\end{equation}
\end{description}

\narrowtext
As we have mentioned before,  only an upper bound of 
$P(\nu_{\mu}\rightarrow \nu_{\tau})$ has been deteermined experimentally. 
So the meaningful inequalities 
arise from the lower bounds of Eqs.(\ref{eq21}) \(\sim\) (\ref{eq24}).
Namely\\
\widetext

{\qquad \qquad \bf Case a}
\begin{equation} 
(s_1c_2-c_1s_2s_3)^2(s_1s_2+c_1c_2s_3)^2\leq
\frac{P(\nu_\mu\rightarrow\nu_\tau)}
{4\sin^2(\frac{\delta
m_{31}^2}{4E_\nu}L)}
 \quad for\quad (s_2^2-c_2^2)(s_1^2-c_1^2s_3^2) \ge 0 \label{eq25}
\end{equation}

{\qquad \qquad \bf Case b}
\begin{equation}
(s_1c_2+c_1s_2s_3)^2(s_1s_2-c_1c_2s_3)^2\leq
\frac{P(\nu_\mu\rightarrow\nu_\tau)}
{4\sin^2(\frac{\delta
m_{31}^2}{4E_\nu}L)} \quad for \quad (s_2^2-c_2^2)(s_1^2-c_1^2s_3^2) \le 0
\label{eq26}
\end{equation}
\narrowtext
From these inequalities (\ref{eq25}) and (\ref{eq26}), 
we obtain the allowed region in the 
$s_1^2$ versus $s_3^2$ plane for a fixed value of $\theta_2$. 
Using $\delta m_{31}^2 \sim \delta m_{21}^2 = 6 
\mbox{eV}^2 \gg \delta m_{32}^2$ 
and \(\frac{P(\nu_\mu \to \nu_\tau)}
{\sin^2 \frac{\delta m_{31}^2 L}{4E_\nu}} < 0.088\), 
we show the allowed regions for 
$\theta_2=0, \frac{\pi}{24}, \frac{\pi}{12},\cdots ,
\frac{\pi}{2}$ in FIG.4.
\begin{center}
--- FIG.4 ---
\end{center}

Next we consider case (ii). Here 
\(\delta m_{31}^2 \sim \delta m_{32}^2 \gg \delta m_{21}^2\), and we have
\begin{equation}
P(\nu_{\mu}\rightarrow\nu_{\tau})=4|U_{\mu 3}|^2|U_{\tau 3}|^2sin^2
(\frac{\delta m_{31}^2L}{4E_\nu}). \label{eq27}
\end{equation}
Here, analogously to case (i), 
\(\sin^2 \left(\frac{\delta m_{21}^2 L}{4 E_\nu}\right) \sim 0\)
has been assumed. Substituting the expression of Eq.(\ref{eq2}) 
into Eq.(\ref{eq27}),
we obtain 
\begin{equation}
P(\nu_\mu \to \nu_\tau) =
4 \sin^2(2.8 \times 10^{-2} m_3^2) s_2^2 c_2^2 c_3^4. \label{eq28}
\end{equation}
Combining Eq.(\ref{eq28}) with the constraints 
from the neutrinoless double 
beta decay Eq.(\ref{eq17}), we also obtain 
\begin{eqnarray}
P(\nu_\mu&\to&\nu_\tau) \label{eq29}\\
   &\ge& 4 \sin^2(2.8 \times 10^{-2} m_3^2) s_2^2 c_2^2 
	\left(\frac{m_3-\langle m_\nu \rangle}{m_3}\right)^2. 
\nonumber
\end{eqnarray}
Using the experimental upper bound of 
\(P(\nu_\mu \to \nu_\tau)<2.5 \times 10^{-3}\) and Eq.(\ref{eq28}), 
we obtain the allowed regions in the \(s_2^2\) versus \(s_3^2\) plane for 
the possible values of \(m_3\) (FIG.5).

\begin{center}
--- FIG.5 ---
\end{center}

Eq.(\ref{eq29}) with \(P(\nu_\mu \to \nu_\tau) < 2.5 \times 10^{-3}\) 
gives more interesting constraints in the \(s_2^2\) versus 
\(\langle m_\nu \rangle\) plane for the possible values of \(m_3\), 
which is depicted in FIG.6.
\begin{center}
--- FIG.6 ---
\end{center}
It should be noted that \(\langle m_\nu \rangle\) can not be larger than 
\(m_3\) as seen from Eq.(\ref{eq16}). 
If we input the experimental upper bound of 
\( \langle m_\nu \rangle \alt O(1\mbox{eV}) \) in FIG.6, 
we obtain more stringent constraints on \(\theta_2\).
It is interesting that the allowed 
region becomes more restrictive as the experimental results become more precise.
That is, the smaller the upper bounds of \(P(\nu_\mu \to \nu_\tau)\) and 
\(\langle m_\nu \rangle\) become, 
the more restricted the allowed region is.
Especially, the large mixing angle solution of \(\theta_2\) becomes 
improbable if \(m_3\) is a candidate for hot dark matter.

\section{Conclusion}

We have obtained constraints from \((\beta \beta )_{0 \nu}\) 
and the CHORUS neutrino oscillation experiment, 
taking into account possible CP violation phases. 
From \((\beta \beta )_{0 \nu}\), 
the allowed regions on \(s_2^2\) versus \(s_3^2\) plane are obtained 
for the respective cases (case(a): \(\langle m_\nu \rangle \le m_1\), 
case(b): \(m_1 \le \langle m_\nu \rangle \le m_2\), 
and case(c): \(m_2 \le \langle m_\nu \rangle \le m_3\) ). 
From the CHORUS neutrino oscillation experiment, 
we have obtained the allowed regions in the \(s_1^2\) versus \(s_3^2\) plane 
for given values of \(\theta_2\) 
for case (i); \(\delta m_{31}^2 \sim 
\delta m_{21}^2 = 6 \mbox{eV}^2 \gg \delta m_{32}^2\). 
For case(ii), 
\(\delta m_{31}^2 \sim \delta m_{32}^2 \gg \delta m_{21}^2\), 
we have obtained the allowed regions 
in the \(s_2^2\) versus \(s_3^2\) plane and 
those in the \(\langle m_\nu \rangle\) versus \(s_2^2\) plane 
for given values of \(m_3\).
Combining the constraint from the CHORUS neutrino oscillation experiment 
with those obtained from \((\beta\beta)_{0 \nu}\), 
we have found that the large mixing angle solution of \(\theta_2\) becomes 
improbable if the neutrino mass \(m_3\) of the third generation 
is a candidate for hot dark matter and not degenerate.

Recently the Super Kamiokande group announced 
that they obtained the large angle 
solution \(\sin^2 2 \theta_{23} \ge 0.8\) with 
\(\delta m_{23}^2=10^{-4} \sim 10^{-2}\)eV\(^2\) 
from the atmospheric neutrino deficit \cite{kami}. 
This conclusion does not contradict our result since their 
\(\delta m_{23}^2\) is of order less than 1eV\(^2\).

\ \\
Acknowledgement:

We are grateful to 
H.Shibuya and Y.Watanabe for informing us of the most recent 
results of the CHORUS group and the Super Kamiokande group, respectively.
We also thank Dr. T. Plewe for careful reading through this paper and 
many useful comments.
\clearpage

\vfill\eject





\clearpage

   \begin{figure}
	\caption{
	Feynman diagram of  neutrinoless double beta decay. }
	\label{fig.1}
   \end{figure}
   \begin{figure}
	\caption{
	The allowed region in the $sin^2\theta_{12}$ versus 
	$\sin^2\theta_{31}$
	plane obtained from neutrinoless double beta decay is 
	given by the shaded areas in the cases: \newline
	(a) $\langle m_\nu \rangle \leq m_1$ \newline
	(b) $m_1 \leq \langle m_\nu\rangle \leq m_2 $ \newline
	(c) $m_2 \leq \langle m_\nu \rangle \leq m_3 $ \newline
	In the allowed region, 
	CP-violation area is specially indicated by 
	the oblique lines.}
	\label{fig.2}
   \end{figure}
   \begin{figure}
	\caption{
	Each allowed region obtained from neutrinoless double beta decay 
	for the cases:\newline
	\(\frac{\langle m_\nu \rangle}{m_3} = 
		0.16,\, 0.33,\, 0.49,\, 0.65,\, 0.82,\, 0.98\)\newline
	under the assumption that 
	\(m_1 \ll \langle m_\nu \rangle < m_2 \sim m_3\) }
	\label{fig.3}
   \end{figure}
    \begin{figure}
	\caption{
	The allowed regions (shaded regions) by 
	the inequalities of 
	Eqs.(3.8) and (3.9) 
	under the condition that
	\(P_{CHORUS}<2.5\times10^{-3}\), 
\(\delta m_{31}^2 \sim \delta m_{21}^2 = 6 
\mbox{eV}^2 \gg \delta m_{32}^2\) 
	with given \(\theta_2\): \newline
	  (a)\(\theta_2=0, \frac{\pi}{2}\)\
	  (b)\(\theta_2=\frac{\pi}{24}, \frac{11\pi}{24}\)\
	  (c)\(\theta_2=\frac{\pi}{12}, \frac{5\pi}{12}\)\
	  (d)\(\theta_2=\frac{\pi}{8}, \frac{3\pi}{8}\) \newline
	  (e)\(\theta_2=\frac{\pi}{6}, \frac{\pi}{3}\) \
	  (f)\(\theta_2=\frac{5\pi}{24}, \frac{7\pi}{24}\) \
	  (g)\(\theta_2=\frac{\pi}{4}\) }
	\label{fig.4}
  \end{figure}
   \begin{figure}
	\caption{
The allowed regions (shaded regions) in the \(s_2^2\) versus \(s_3^2\) 
	plane
for the possible values of \(m_3\) obtained from the CHORUS experiment.}
	\label{fig.5}
   \end{figure}
   \begin{figure}
	\caption{
	The allowed regions (shaded regions) 
	in the \(\langle m_\nu \rangle\) versus \(s_2^2\) plane 
	for the possible values of \(m_3\) obtained from the CHORUS 
	experiment and \(( \beta \beta)_{0 \nu}\). }
	\label{fig.6}
   \end{figure}

\clearpage

 \begin{figure}[htbp]
 	\begin{center}
 	\leavevmode
 	\epsfile{file=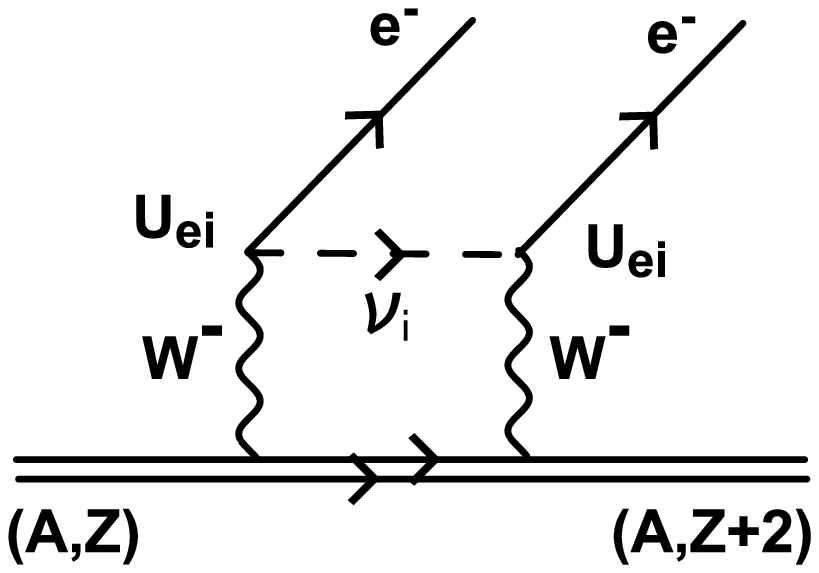,width=8.6cm}\\
 \ \\
 	{\Huge FIG.1}
 	\end{center}
 \end{figure}

 \begin{figure}[hbtp]
 	\begin{center}
 	\leavevmode
 	\epsfile{file=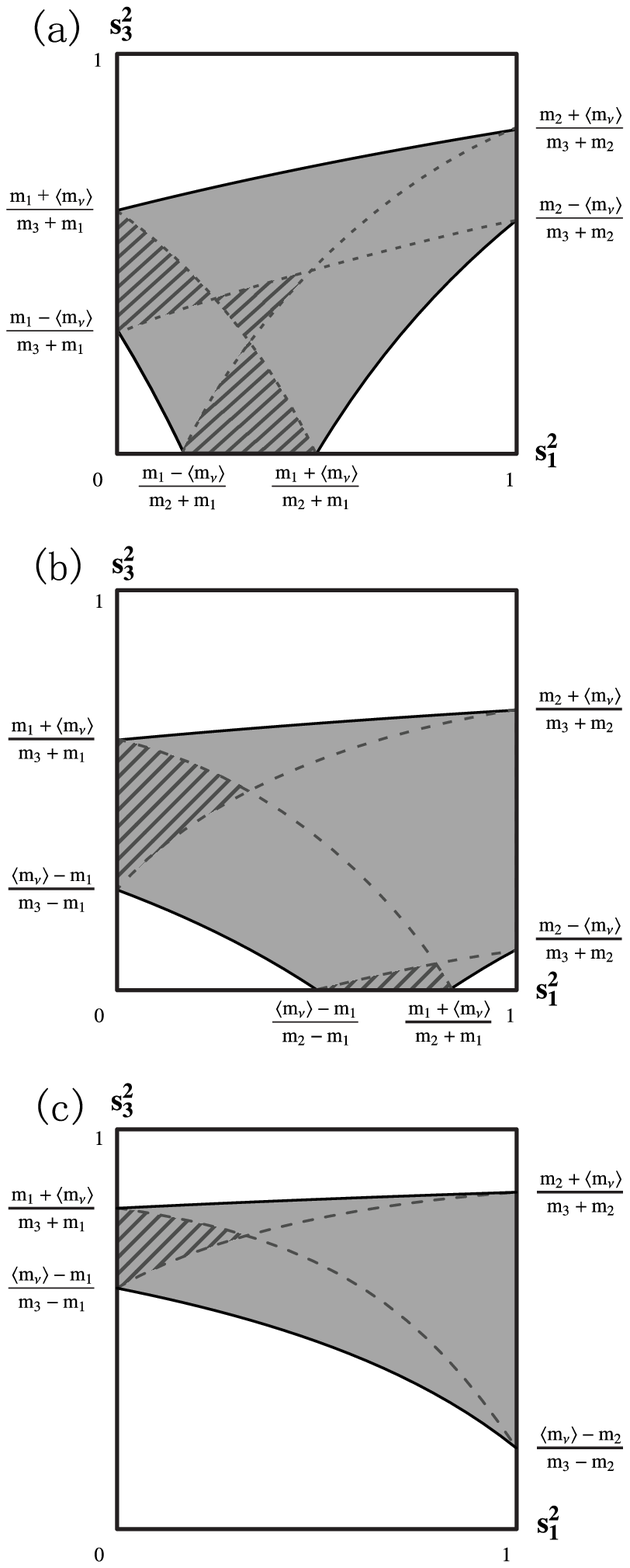,width=8.6cm}\\
 \ \\
 	{\Huge FIG.2}
 	\end{center}
 \end{figure}

 \begin{figure}[htbp]
 	\begin{center}
 	\leavevmode
 	\epsfile{file=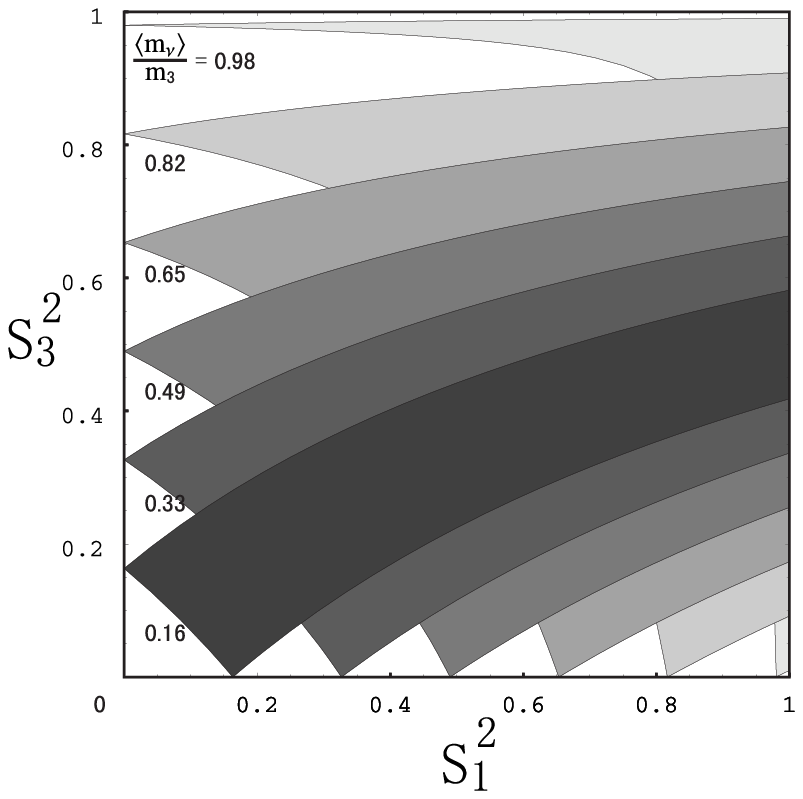,width=8.6cm}\\
 \ \\
 	{\Huge FIG.3}
 	\end{center}
 \end{figure}

\widetext
 \begin{figure}[htb]
 	\begin{center}
 	\leavevmode
 	\epsfile{file=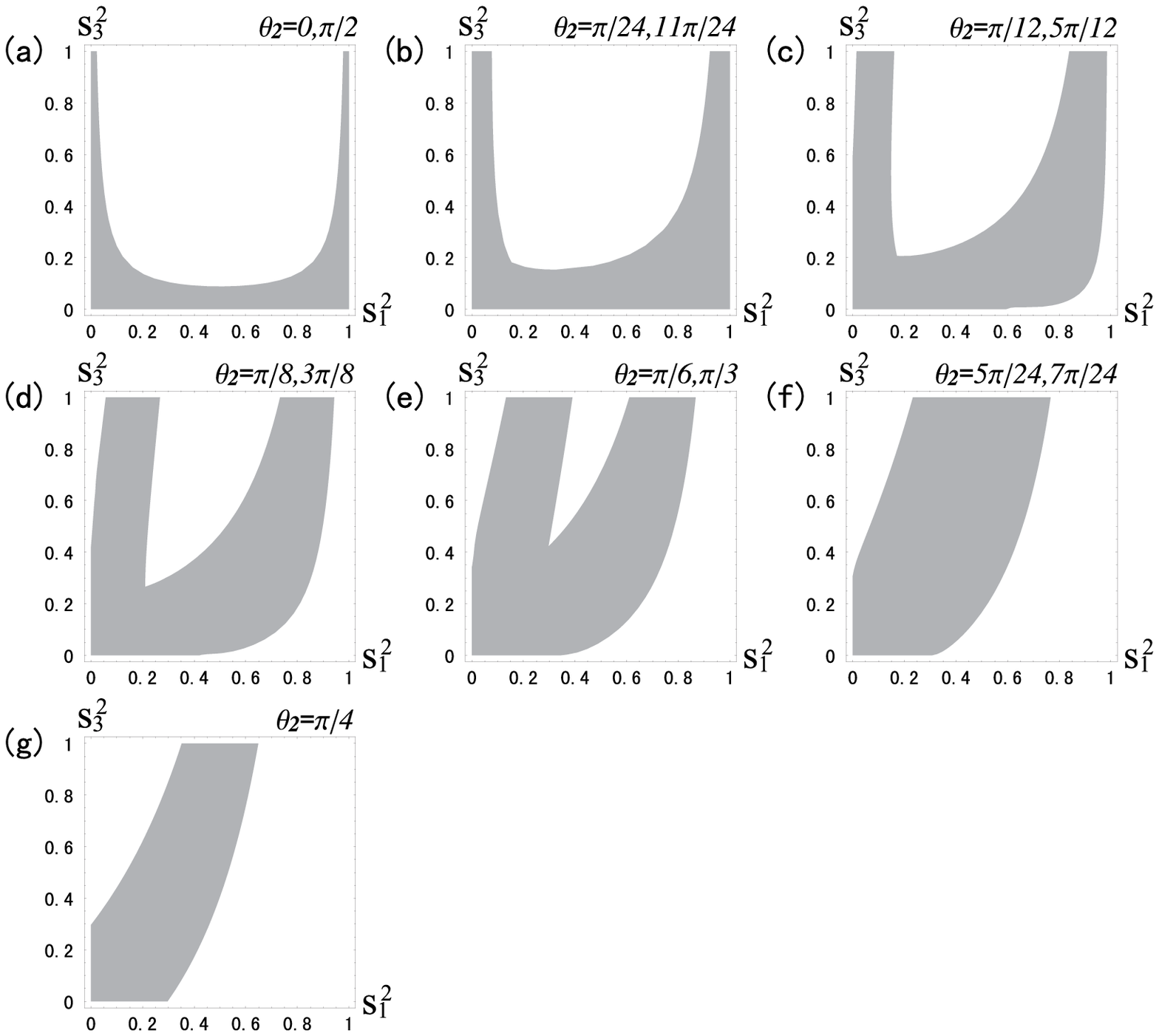,width=17.2cm}\\
 \ \\
 	{\Huge FIG.4}
 	\end{center}
 \end{figure}

\clearpage
 \begin{figure}[htb]
 	\begin{center}
 	\leavevmode
 	\epsfile{file=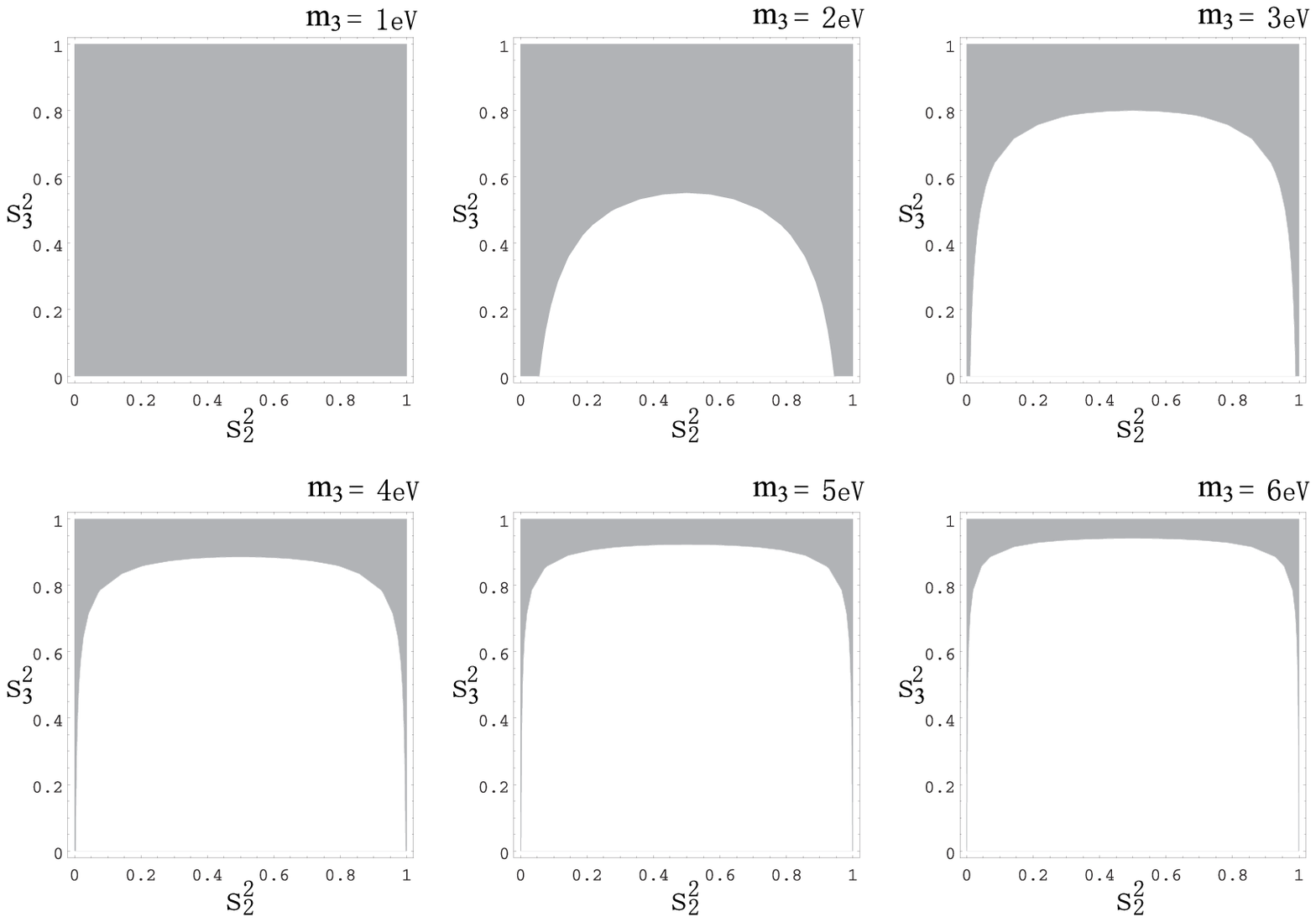,width=17.2cm}\\
 \ \\
 	{\Huge FIG.5}
 	\end{center}
 \end{figure}

 \begin{figure}[htb]
 	\begin{center}
 	\leavevmode
 	\epsfile{file=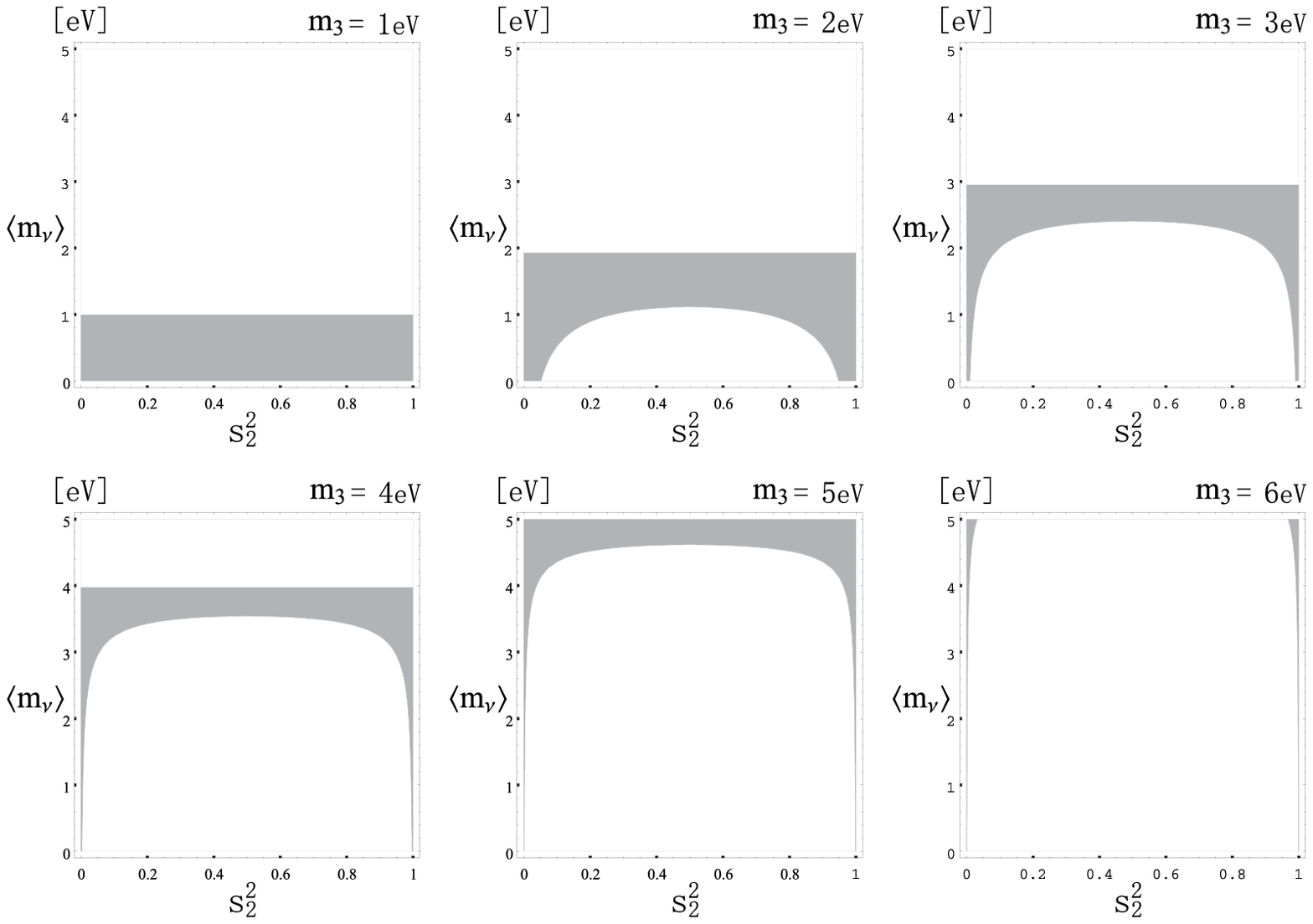,width=17.2cm}\\
 \ \\
 	{\Huge FIG.6}
 	\end{center}
 \end{figure}

\end{document}